\documentclass{article}
\usepackage[utf8]{inputenc}

\title{Dynamic data fusion using multi-input models for malware classification}
\author{Viktor Zenkov$^1$, Jason Laska$^2$}
\date{July 2019}

\usepackage{graphicx}
\usepackage{amsthm}
\usepackage{amsmath} 
\usepackage[table]{xcolor} 

\usepackage{url} 
 \usepackage{multirow}

\def\rot{\rotatebox}   

\begin{document}

\maketitle

$^1$Department of Computer Science and Electrical Engineering, University of Tennessee, Knoxville

$^2$Cyber and Information Security Research Group, Oak Ridge National Laboratory

\section{Introduction}

In the 21st century, we have seen a trend of common systems, such as vehicles, routers, and light bulbs, becoming internet-connected \emph{cyber-systems}. This means that these systems are connected to the internet or are in regular contact with the internet or other cyber-systems, making the cyber-system potentially vulnerable to corruption, via \emph{malware}. Malware is code written by cybercriminals with the intent of hijacking cyber-systems and causing unwanted consequences. To help counteract the growing dangers of malware, it will be useful to quickly classify different forms of malware, making it easier to identify attacks and build defenses against future attacks. Here we utilized \emph{machine learning} to train a computer to classify malware.

We used malware data provided by Microsoft \cite{microsoft} and constructed four algorithms to classify malware log files. We split the data into two data streams, which we call text and hex commands, which we used as the bases for two of our algorithms, which we call \emph{text} and \emph{hex}. This is shown in Figure \ref{preprocessing}. Also, we fused the data in two ways to form two more algorithms, which we call \emph{multi-input} and \emph{ensemble}. We trained neural networks on a high-performance computer using the four algorithms and compared how well the resulting models classified our test data. We found that the multi-input model performed best.

\section{Related Work}

In order to analyze malware, we need to turn the malware files into data that can be used as inputs to machine learning. Malware is sometimes classified using \textbf{opcodes}, which are machine instructions in hexadecimal format. Opcodes are used by Sewak et al \cite{SewakMohit2018CoDL} to compare different neural network algorithms. In our paper we also used opcodes for classifying, and additionally we used text data consisting of metadata extracted from the opcodes.

Two popular ways to analyze malware files are called dynamic and static program analysis. \textbf{Dynamic} program analysis involves executing  malware files and observing their behavior in real time, described in Zhao et al \cite{ZhaoEtAl}. \textbf{Static} program analysis involves analyzing the structure of malware files without running the malicious software.  An alternative to classifying malware ``after the fact'' is detecting malware as it is enacted in real time, which can be done using a visual analytics system like Situ, discussed in Goodall et al \cite{GoodallJohnR2019SIaE}. Here we use static program analysis.

The Microsoft malware dataset we use has been used by others employing neural networks to classify malware, such as Chen \cite{ChenLiMal}, who used transfer learning to transform malware into images to perform  classification. Chen used Microsoft's malware dataset as one of three datasets to test the neural network; using three datasets helped give confidence in the neural network's effectiveness.  Chen achieved an accuracy of 98.1 on the Microsoft data, though with performance measures that did not explicitly account for the imbalance of the Microsoft dataset. We account for that imbalance in this paper.

Another previous use of Microsoft's malware dataset was in Raff et al \cite{RaffEtAl}, who like us used static program analysis. Raff used n-grams with n = 4, 5, and 6 for machine learning, but found that the performance of n-grams was worse than expected. Raff speculated that the neural network may have sometimes learned to classify executables instead of malware.

\section {Method}

To train a computer to classify malware, we built a \textbf{neural network}, which is a multi-level network of nodes which processes labeled data to form a \textbf{model}. We can then use the model to analyze an unlabeled malware file and calculate a likely classification for the file.

\subsection{Data}

To train a neural network and create a model to classify malware, we needed a dataset of malware files where each file already had a label corresponding to a family (classification) of malware. Microsoft provided an open-source dataset of disassembled malware for a Kaggle challenge \cite{microsoft}. The disassembly process resulted in assembly language source code (asm) files; an excerpt is shown in Figure \ref{preprocessing}. The dataset included 10,000 training files, with one of nine possible labels assigned to each file. The nine families of malware are called Ramnit (1,541 files), Lollipop (2,478 files), Kelihos\_ver3 (2,942 files), Vundo (475 files), Simda (42 files), Tracur (751 files), Kelihos\_ver1 (398 files), Obfuscator.ACY (1,228 files), and Gatak (1,013 files). Each label corresponded to a family of malware. As can be seen, the files are not balanced evenly between the labels; for example, Label Kelihos\_ver3 is represented 70 times as much as Label Simda. This can skew our metrics for assessing the effectiveness of our neural networks to downplay if a neural network ignores an uncommon label. We therefore include metrics that take into account the imbalance.

\subsection{Data Streams}
 
We used the Python package Keras \cite{chollet2015keras}, a neural networks Application Programming Interface, to write code to train neural networks on the data. First we needed to filter the files into sequences of specific elements which could be easily quantified and processed. We called the elements \textbf{text commands} and \textbf{hex commands}. 

\begin{figure}
\begin{center}
\includegraphics[width=0.8\textwidth]{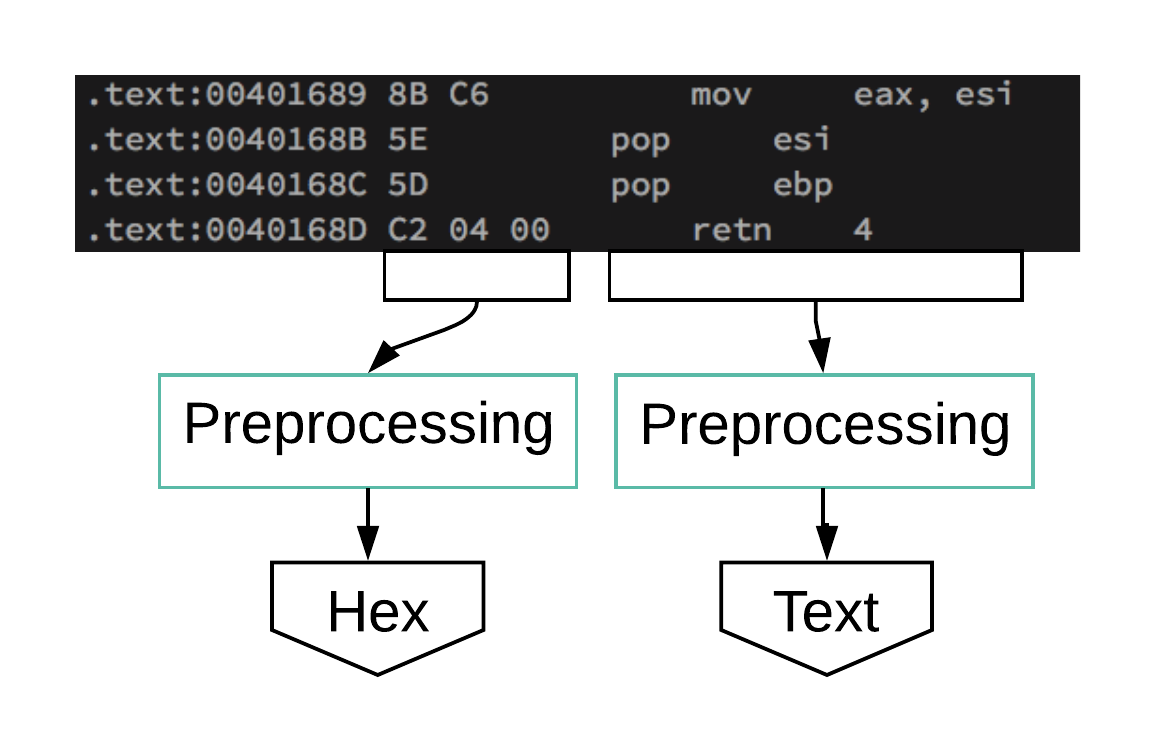}
\caption{An excerpt from an assembly language source code  file. First we gather text and hex commands (shown here), then we transform them into integers.\label{preprocessing}}%
\end{center}
\end{figure}

To get the filtered files, we \emph{parsed} the files using the Python package Pyparsing \cite{McGuire:2007:GSP:1406599}.  For hex commands, we gathered the hexadecimal series of numbers from each line where we acquired the text commands. This is shown in the left of Figure \ref{preprocessing}. For text commands, we gathered the Assembly code in each line which occurs after hexadecimal series of numbers and before comments (the comments are not shown in the example). This is shown in the right of Figure \ref{preprocessing}. After the parsing process, we possessed, for each original asm file, two files consisting of text commands or hex commands.

\subsection{Preprocessing}

The framework we used to implement machine learning requires the datapoints, which are each malware files, to be lists of integers so the data can later be transformed into vectors. Therefore, we needed to translate our text and hex commands into lists of integers. For text and hex files separately, we thus replaced each unique space-separated \textbf{word} with an assigned integer. The words were encoded such that high frequency values received smaller integer values. There were over 80 million unique text commands, in part due to sequences of text commands including positions in memory, of which there are many possibilities. We had roughly 256 unique hex commands because all of the hex commands are two-digit hexadecimal numbers, and there are only 256 possible two-digit hexadecimal numbers.

At this stage, we have two sets of files. All the files are made up of integers; one set contains integers from text commands, while the other set contain integers from hex commands. We will refer to the files as text and hex files from now on.

Next we needed to shrink the data to a manageable size to ensure that training finished within 24 hours. This was necessary to enable us to train and test many neural networks with different parameters over a period of months. After the previously mentioned preprocessing, the text and hex training files had an average length of 80,000 and in the text files' case contained 80 million unique integers. These sizes required tens of hours to train, so we needed to shorten the files in order to have time to train multiple times. To contain the runtime into less than 24 hours per run, we kept the 5,000 most commonly occurring text integers and shortened the text and hex files to a length of at most 15,000 integers.

\subsection {Neural Networks}

Neural networks train \emph{parameters} which are used for calculating the likelihoods of each datapoint falling under each possible classification. The neural network does this repeatedly; each loop through the data is called an \emph{epoch}. In each epoch, the neural network optimizes the parameters by calculating gradients. At the end of each epoch, an \textbf{accuracy} is calculated. In this case, the accuracy is the mathematical fraction 
\begin{equation}
\frac{\text{number of files with labels correctly predicted}}{\text{total number of files}}.
\end{equation} 
We separated our data into main training data (75\%), validation data (15\%, used to evaluate generalizability), and test data (10\%, used for calculating the final accuracy). The accuracy at the end of the epoch is calculated based on the validation data and is used to tune the parameters. To ensure that our data was separated consistently, we seeded both the Python and the TensorFlow random functions.

To ensure that our neural network finished training in a reasonable amount of time and avoided \emph{overfitting}, in which the neural network becomes too attuned to the training data and ceases to be generalizable and useful for the validation and test data, we allowed training for many epochs but  stopped early if an epoch failed to increase the validation accuracy by more than 0.005. 

We designed four algorithms with which to train neural networks. Two of them were  single-input algorithms, taking in only the \textbf{text} or \textbf{hex} integer data that resulted from our preprocessing stage. The other two are combinations: the \textbf{multi-input} algorithm concatenates the two single-input algorithms partway through the neural network building process, while the \textbf{ensemble} algorithm classifies based on combining the probabilities output by the single-input algorithms.  This paper compares these four algorithms' effectiveness in classifying the malware data. \footnote{\raggedright The code used to create the models is hosted at https://github.com/viktorZenkov/MalwareClassification.}

\subsection {Single-Input Algorithms}

\begin{figure}
\begin{center}
\includegraphics[width=0.7\textwidth]{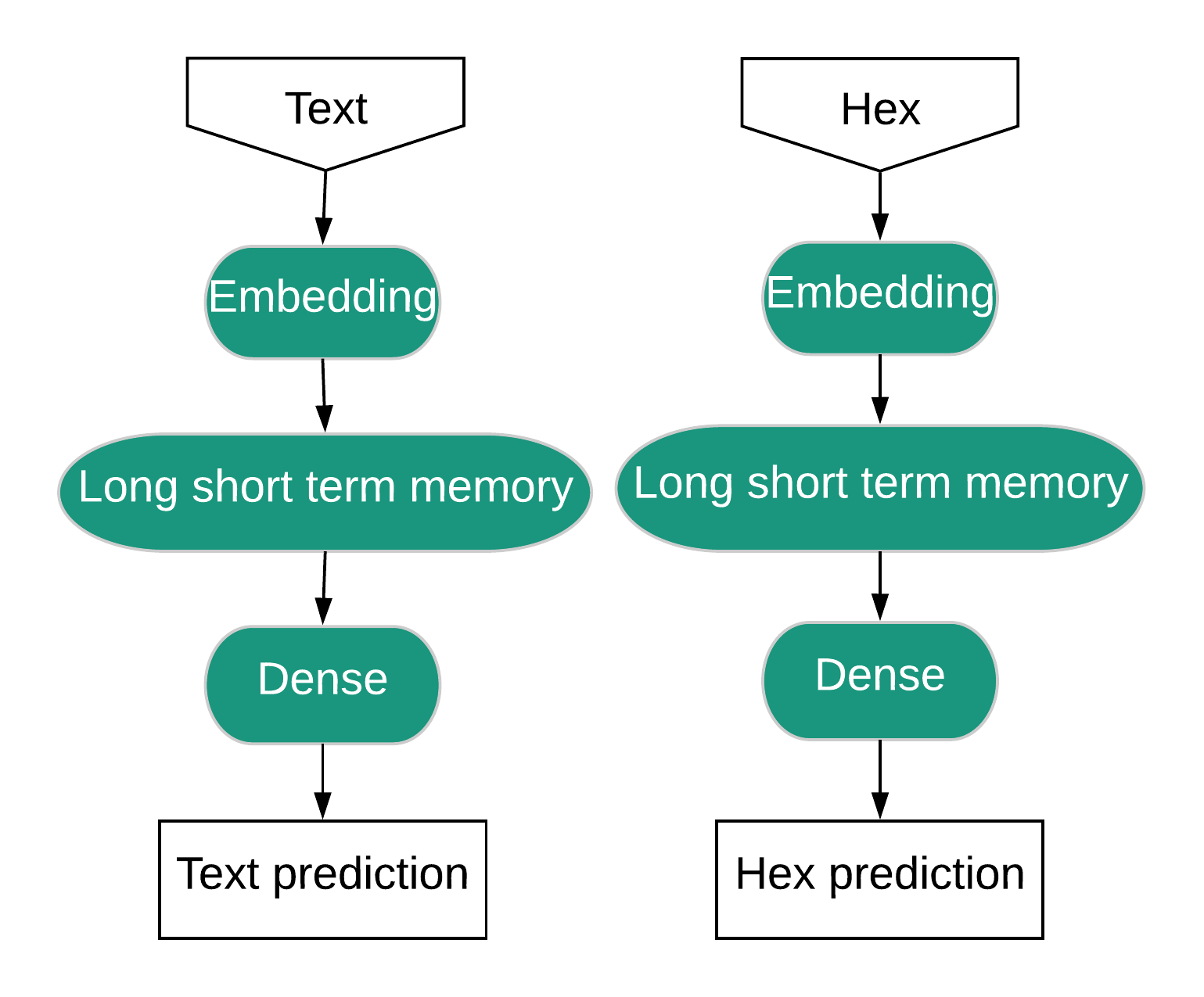}
\caption{Single-Input Algorithms use Text or Hex data to make predictions. \label{architSingle}}%
\end{center}
\end{figure}

For both the text and the hex files, we used a \emph{layered architecture}, which is a sequence of layers which process the data. We started with an \emph{embedding} layer which replaces each integer with a dense 128-dimensional vector. The choice of 128 dimensions was chosen by trial and error of powers of 2. Then we added a \emph{long short term memory} (LSTM) layer, which learns a representation of sequential data \cite{lstm1997}. Finally, we added a \emph{dense layer}, which uses \emph{softmax}, a function that compresses data to form \emph{probabilities} that a datapoint will be each of the nine possible classes. These algorithms are shown in Figure \ref{architSingle}.


\subsection {Multi-Input Algorithm}

\begin{figure}
\begin{center}
\includegraphics[width=0.7\textwidth]{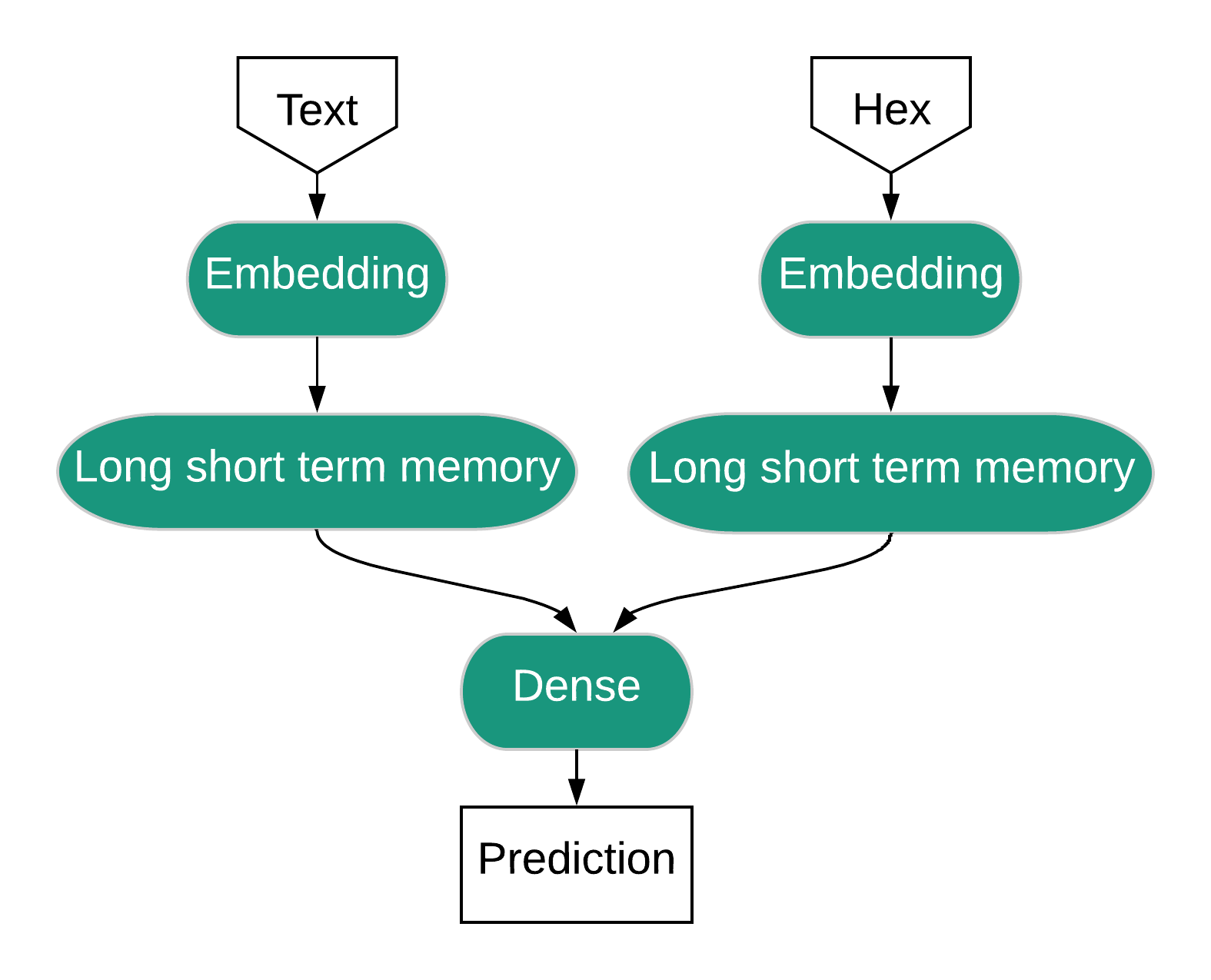}
\caption{The Multi-Input Algorithm uses both Text and Hex data to make predictions. \label{architMulti}}%
\end{center}
\end{figure}

The multi-input algorithm also used a layered architecture. We started with embedding and LSTM layers for text and hex, as before. We then concatenated the outputs of the two LSTM layers, sending the output of each layer to one new layer.  We added a dense layer with softmax after the concatenation. Finally, we fit the neural network using both the text and hex data. This algorithm is shown in Figure \ref{architMulti}.

\subsection {Ensemble Algorithm}

\begin{figure}
\begin{center}
\includegraphics[width=0.7\textwidth]{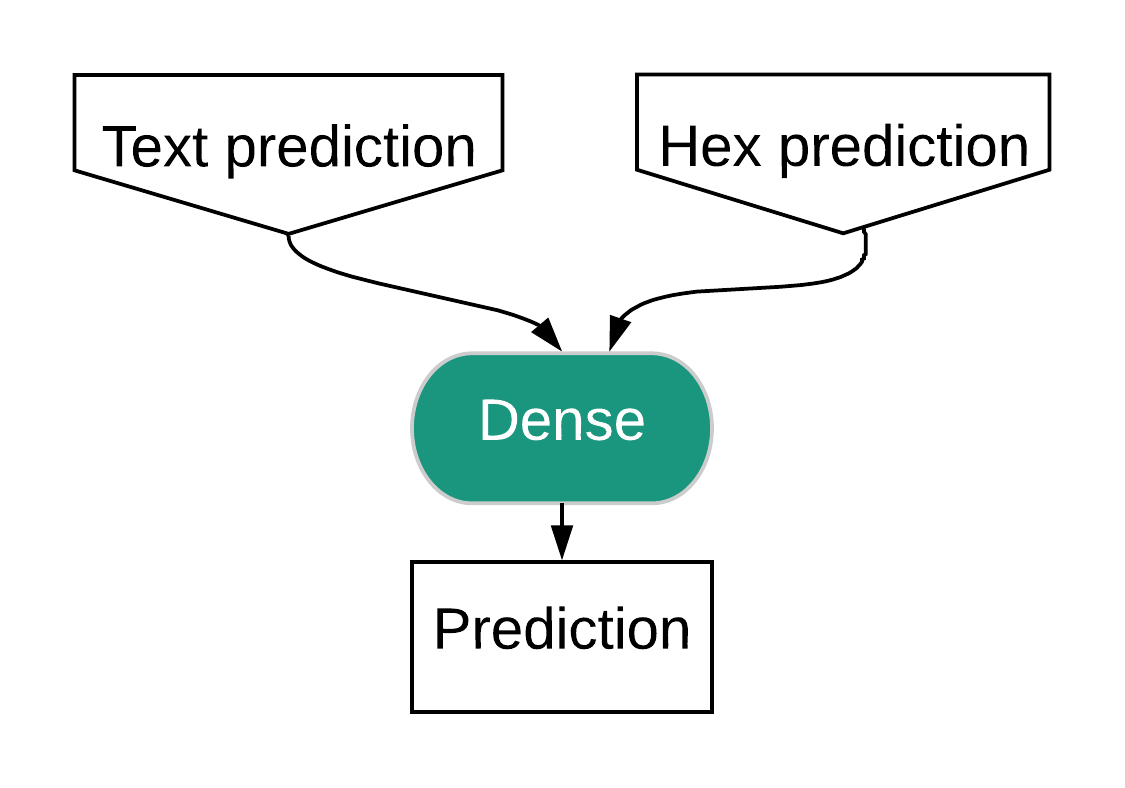}
\caption{The Ensemble Algorithm uses predictions of both single-input algorithms to make predictions.\label{architEnsemble}}%
\end{center}
\end{figure}

For the ensemble algorithm, we first acquired the predicted classification probabilities for all the files using the text and hex formats from the Single-Input algorithms. This data was in the form of two arrays containing sets of probabilities for classes for each file.  Then we concatenated these probability arrays to obtain one array, and each datapoint's entry became this concatenated array of two sets of probabilities. We added a dense layer with softmax. Finally, we fit the neural network using these probabilities from the text and hex. This algorithm is shown in Figure \ref{architEnsemble}. 

\subsection {Computing Resources}

We ran our training for up to 24 hours at a time on a high performance computer, using similar nodes each run. This time limit gave a clear benchmark for tuning neural network parameters.

\subsection{Training}

We varied parameters to train neural networks, searching for the best set of parameters for each algorithm. For the text, hex, and multi-input algorithms, we varied the following parameters:  patience (the number of  epochs without an improvement after which training is terminated, where an improvement is an increase of at least 0.005 in the validation accuracy), dropout and recurrent dropout rate (the fraction of data to drop for the transformation of the inputs and the recurrent state),  dimensionality of the output of the LSTM, max features (number of unique integers),   max length (number of integers per file),   percent of the data to use as validation data, and batch size (the amount of data processed at a time before updating the neural network's weights) \cite{chollet2015keras}. For the ensemble algorithm, the only parameters were the test accuracies (defined below) of the text and hex algorithm used to make the ensemble algorithm. 

For each combination of parameters for each algorithm, we trained ten times and calculated the test accuracy for each training run. In each row of each table we list the mean and maximum test accuracy for the ten runs. For the neural network with the highest test accuracy between all the combinations of parameters, we  provide the (normalized) \textbf{confusion matrix}. A confusion matrix is a matrix whose rows and the columns correspond to the classes in the same order as listed in the Data section, with rows signifying the real class and columns signifying the predicted class, and whose contents are the \emph{counts} of datapoints which are in each real class and which were predicted as being in each  class.

In the following subsections, we provide the \emph{normalized} confusion matrix (which has \emph{proportions} in each cell instead of the count, with each row summing to 1) for the neural network  that had the largest test accuracy  from each algorithm. Additionally, we provide the values of three measures of performance:
\begin{itemize}
\item The test accuracy of the neural network
\item The balanced accuracy of the neural network
\item The average table of confusion accuracy
\end{itemize}

The \textbf{test accuracy} is the sum of the diagonal elements of a (non-normalized) confusion matrix (the overall count of  correct predictions) divided by the sum of all the elements (counts) in the confusion matrix. For each of our combinations of parameters for each algorithm, we provide the average and the maximum of the ten runs' test accuracies in the Text, Hex, and Multi-Input subsections, and we provide the maximum of the ten runs' test accuracies in the Ensemble subsection.

The \textbf{balanced accuracy} is  the average of the diagonal elements in the normalized confusion matrices. This takes into account any classes which have few data points and which are mostly ignored by the  test accuracy. 

\begin{figure}
\begin{center}
\includegraphics[width=1.0\textwidth]{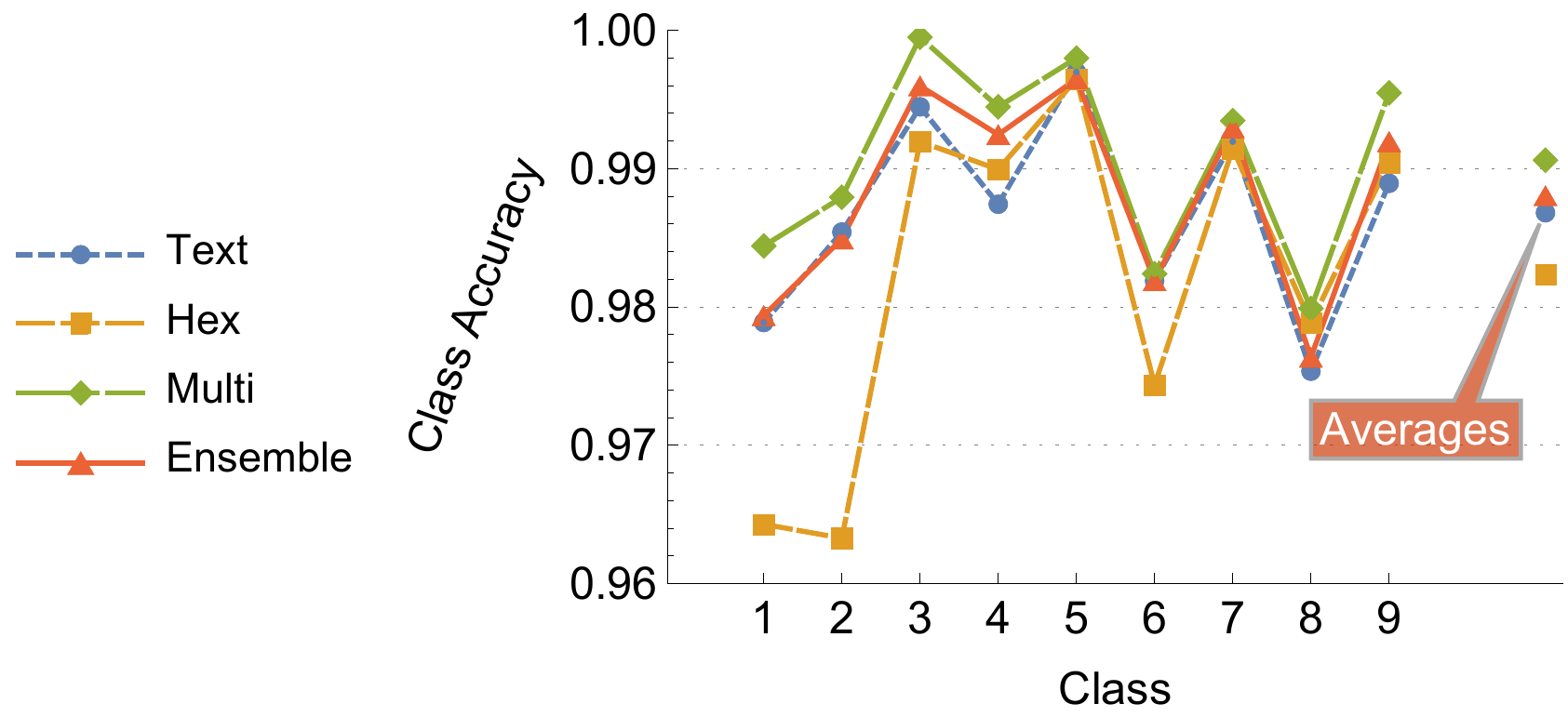}
\caption{For each class (as ordered in the Data section), the average table of confusion accuracy for the best model of each algorithm is shown. The multi-input model performed best for all nine classes and the hex model performed worst for seven classes. The test and ensemble models performed similarly for most classes, with ensemble typically performing slightly better than text. These trends are reflected in the average test accuracies. \label{classAccuracies}}%
\end{center}
\end{figure}

A \textbf{table of confusion} for a particular class is a 2-by-2 table classifying all data as belonging to the class or not belonging to the class; as this is a confusion matrix, its test accuracy may be calculated as described above. The \textbf{average table of confusion accuracy} is the average of the test accuracies of the tables of confusion for all the classes. The average table of confusion accuracy is typically large due to the large quantity of true negatives. The table of confusion accuracies and their averages for each algorithm are shown in Figure \ref{classAccuracies}.

\section{Results}

\subsection{Text Algorithm Results}

The text algorithm performed third best of the four algorithms. Its runtime was not significantly shorter than the best algorithm, but it has less required preprocessing. The best neural network produced a test accuracy of 0.94, a balanced accuracy of 0.83, and an average table of confusion accuracy of 0.99. (The patience was 20, the dropout and recurrent dropout were 0.05, the dimensionality of the LSTM output was 150, the maximum number of unique integers was 5,000, the maximum number of integers per file was 15,000,  the validation split was 0.15, and the batch size was 32. This is shown in the last row of Table \ref{textTable}.)

\begin{table}
\centering
\begin{tabular}{|c|c|c|c|c|c|c||c|c|}
\hline
\rot{75}{patience} & \rot{75}{dropout} & \rot{75}{dimensionality} & \rot{75}{max features} & \rot{75}{max length} & \rot{75}{validation split} & \rot{75}{batch size} & \rot{75}{avg. accuracy} & \rot{75}{max accuracy}  \\
\hline
\hline 1 & 0.1 & 150 & 5,000 & 15,000 & 0.15 & 16 & 0.82 & 0.88 \\
\hline 2 & 0.1 & 150 & 5,000 & 15,000 & 0.15 & 16 & 0.88 & 0.89 \\
\hline 3 & 0.1 & 150 & 5,000 & 15,000 & 0.15 & 16 & 0.79 & 0.89 \\
\hline 3 & 0.2 & 150 & 5,000 & 15,000 & 0.15 & 16 & 0.82 & 0.93 \\
\hline 3 & 0.0 & 150 & 5,000 & 15,000 & 0.15 & 16 & 0.85 & 0.90 \\
\hline 3 & 0.05 & 256 & 5,000 & 15,000 & 0.15 & 16 & $<0.7$ & $<0.7$ \\
\hline 3 & 0.05 & 150 & 10,000 & 15,000 & 0.15 & 16 & 0.86 & 0.90 \\
\hline 3 & 0.05 & 150 & 5,000 & 25,000 & 0.15 & 16 & 0.84 & 0.89 \\
\hline 3 & 0.05 & 150 & 5,000 & 15,000 & 0.3 & 16 & 0.80 & 0.89 \\
\hline 3 & 0.05 & 150 & 5,000 & 15,000 & 0.15 & 32 & 0.82 & 0.88 \\
\hline 4 & 0.05 & 150 & 5,000 & 15,000 & 0.15 & 32 & 0.86 & 0.89 \\
\hline 6 & 0.05 & 150 & 5,000 & 15,000 & 0.15 & 32 & 0.87 & 0.89 \\
\hline 20 & 0.05 & 150 & 5,000 & 15,000 & 0.15 & 32 & 0.9 & \cellcolor{lightgray} 0.94 \\ 
\hline
\end{tabular}
\caption{Text algorithm: Each row shows the maximum and mean test accuracies for ten runs. The best neural network from all the combinations of parameters has a test accuracy of 0.94. Notable aspects of the best combination of parameters include the higher patience and batch size than many of the other combinations.}
\label{textTable}
\end{table}

The normalized confusion matrix in Table \ref{tablesomething} results from the best neural network (test accuracy of 0.94). The highlighted diagonal of the text confusion matrix shows the proportions of predictions for each category that were correct. This best neural network  predicted correctly above 80\% on all classes except class 5, which was only predicted correctly 15\% of the time. However, the neural network never incorrectly predicted a different class to be class 5.

In conclusion, the text model performed third best, better than the other single input model (hex) and worse than the fused models (multi-input  and ensemble).

\begin{table}
\centering
\makebox[\textwidth]{
\begin{tabular}{c|c|c|c|c|c|c|c|c|c|}  
\multicolumn{10}{c}{Predicted Class}\\  
\cline{2-10}  
 \multirow{9}{*}{\rotatebox[origin=c]{90}{Real Class}} & \cellcolor{lightgray} 0.92 & 0.0182 & 0.0109 & 0.0036 & 0.  &  0.0255 & 0.   &  0.0145 & 0.0073\\  
\cline{2-10}   
 & 0.0040 & \cellcolor{lightgray} 0.9737 & 0.  &  0.0020 & 0.  & 0.0101 & 0.0020  & 0.0081 & 0. \\
\cline{2-10}  
 & 0.0019  &  0.   &  \cellcolor{lightgray}   0.9963 &  0.   &  0. &   0.0019  &  0. &     0.     &      0. \\
\cline{2-10}  
 & 0. & 0.0118 &  0.0235 &  \cellcolor{lightgray} 0.8824 & 0.  & 0.0588   & 0. &   0.0118 &  0.0118 \\
\cline{2-10}  
 & 0. & 0.2857  & 0. & 0.1429  & \cellcolor{lightgray}0.1429  & 0.2857  &  0. & 0.1429 &  0. \\
\cline{2-10}  
 & 0. & 0. & 0.0148 & 0. & 0. & \cellcolor{lightgray} 0.9703 & 0.0074 & 0.0074 & 0. \\
\cline{2-10} 
 & 0.0694 & 0.0556 & 0. & 0. & 0. & 0.0278 & \cellcolor{lightgray} 0.8194 & 0.02778 & 0.        \\
\cline{2-10} 
 & 0.0585 & 0.0049 & 0.0098 & 0.0195 & 0. & 0.0439 & 0.0049 & \cellcolor{lightgray} 0.8390 & 0.0195  \\
\cline{2-10}  
 & 0. & 0.0169 & 0. & 0.0449 & 0. & 0.0056 & 0. & 0.0168 & \cellcolor{lightgray} 0.9157 \\
\cline{2-10} 
\end{tabular}}
\caption{Text algorithm: normalized confusion matrix for neural network with accuracy of 0.94. The balanced accuracy was 0.83 and the average table of confusion accuracy was 0.99 (not shown in the table).}
\label{tablesomething}
\end{table}

\subsection{Hex}

The hex algorithm performed worst of the four algorithms.  The best neural network produced a best test accuracy of 0.92, a balanced accuracy of 0.80, and an average table of confusion accuracy of 0.98. (The patience was 20, the dropout and recurrent dropout were 0.05, the dimensionality of the LSTM output was 150, the maximum number of unique integers was 5,000, the maximum number of integers per file was 15,000,  the validation split was 0.15, and the batch size was 32. These are the same parameters that resulted in the best test accuracy for the text data. This is shown in the last row of Table \ref{table55}.)


\begin{table}
\centering
\begin{tabular}{|c|c|c|c|c|c|c||c|c|}
\hline
\rot{75}{patience} & \rot{75}{dropout} & \rot{75}{dimensionality} & \rot{75}{max features} & \rot{75}{max length} & \rot{75}{validation split} & \rot{75}{batch size} & \rot{75}{avg. accuracy} & \rot{75}{max accuracy}  \\
\hline
\hline 6 & 0.05 & 64 & 5,000 & 15,000 & 0.15 & 32 & 0.67 & 0.80 \\
\hline 6 & 0.05 & 150 & 5,000 & 15,000 & 0.15 & 32 & 0.66 & 0.88 \\
\hline 20 & 0.05 & 150 & 5,000 & 15,000 & 0.15 & 32 & 0.88 & \cellcolor{lightgray} 0.92 \\ 
\hline 28 & 0.05 & 150 & 5,000 & 15,000 & 0.15 & 32 & 0.80 & 0.92 \\
\hline
\end{tabular}
\caption{Hex algorithm: Each row shows the maximum and mean test accuracies for ten runs. The best neural network from all the combinations of parameters has a test accuracy of 0.92. We started our training using the best text neural network parameters, and in fact these are the same parameters that resulted in the best neural network for the hex algorithm.}
\label{table55}
\end{table}

The normalized confusion matrix in Table \ref{table44} results from the best neural network  (test accuracy of 0.92). The highlighted diagonal of the hex confusion matrix shows the proportions of predictions for each category that were correct. This best neural network  predicted correctly above 75\% on all classes except class 5, which the neural network never predicted.

In conclusion, the hex model performed worst. Despite having the same training parameters as the best text model, the model based on hex data alone could not predict as well.

\begin{table}
\centering
\makebox[\textwidth]{
\begin{tabular}{c|c|c|c|c|c|c|c|c|c|}
\multicolumn{10}{c}{Predicted Class} \\ 
\cline{2-10}  \multirow{9}{*}{\rotatebox[origin=c]{90}{Real Class}}  & \cellcolor{lightgray} 0.8618  & 0.0691 &  0.0036  & 0.& 0. &    0.0327  & 0.0109  & 0.0182  & 0.0036 \\
\cline{2-10} &  0.0182  & \cellcolor{lightgray} 0.9332  & 0.0081  & 0.0121  & 0. & 0.0081 &   0.0040  & 0.0101  & 0.0061 \\
\cline{2-10} &  0.0037  & 0. & \cellcolor{lightgray} 0.9870  & 0.0037  & 0. & 0.0056  &  0.      &     0.       &    0. \\
\cline{2-10}  & 0.0118  & 0. & 0.0118  &  \cellcolor{lightgray} 0.8588  & 0. & 0.1059 &   0. &     0.0118 &  0. \\
\cline{2-10}  & 0. &   0.2857  & 0.      &     0. &     \cellcolor{lightgray} 0.       &    0.2857   & 0.    &       0. &    0.4286 \\
\cline{2-10} &  0.0074  & 0.0296 &  0.0148 & 0.   & 0. &    \cellcolor{lightgray} 0.9407 &   0. &    0.0074 &  0. \\ 
\cline{2-10}  & 0.0417  & 0.0833  & 0.    & 0. &     0. &  0.0278   &  \cellcolor{lightgray}0.8333  & 0.0139  & 0.    \\     
\cline{2-10}  & 0.0585  & 0.0293  & 0.0049  & 0.     &      0. & 0.0488   & 0. & \cellcolor{lightgray} 0.8585  & 0. \\        
\cline{2-10}  & 0.0281  & 0.0169  & 0.    & 0. &    0. &    0.0225  &  0. &    0. &    \cellcolor{lightgray} 0.9326 \\
\cline{2-10}
\end{tabular}}
\caption{Hex algorithm: normalized confusion matrix for neural network with accuracy of 0.92. The balanced accuracy was 0.80 and the average table of confusion accuracy was 0.98 (not shown in the table).}
\label{table44}
\end{table}

\subsection{Multi-Input}

The multi-input algorithm performed best of the four algorithms. The best neural network produced a best test accuracy of 0.96, a balanced accuracy of 0.90, and an average table of confusion accuracy of 0.99. (The patience was 50, the dropout and recurrent dropout were 0.1, the dimensionality of the LSTM output was 150, the maximum number of unique integers was 5,000, the maximum number of integers per file was 15,000,  the validation split was 0.15, and the batch size was 32. This is shown in the last row of Table \ref{table88}.)


\begin{table}
\centering
\begin{tabular}{|c|c|c|c|c|c|c||c|c|}
\hline
\rot{75}{patience} & \rot{75}{dropout} & \rot{75}{dimensionality} & \rot{75}{max features} & \rot{75}{max length} & \rot{75}{validation split} & \rot{75}{batch size} & \rot{75}{avg. accuracy} & \rot{75}{max accuracy}  \\
\hline
\hline 10 & 0.1 & 150 & 5,000 & 15,000 & 0.15 & 32 & 0.84 & 0.89 \\
\hline 28 & 0.1 & 150 & 5,000 & 15,000 & 0.15 & 32 & 0.90 & 0.95 \\
\hline 50 & 0.1 & 150 & 5,000 & 15,000 & 0.15 & 32 & 0.93 &  \cellcolor{lightgray} 0.96 \\ 
\hline
\end{tabular}
\caption{Multi-input algorithm: Each row shows the maximum and mean test accuracies for ten runs. The best neural network from all the combinations of parameters has a test accuracy of 0.96.}
\label{table88}
\end{table}

The normalized confusion matrix in Table \ref{table99} results from the best neural network (test accuracy of 0.96). The highlighted diagonal of the multi-input confusion matrix shows the proportions of predictions for each category that were correct. This multi-input neural network  predicted correctly above 85\% on all classes except class 5, though it  predicted correctly on class 5 over  half of the time, which was much more than the  other three algorithms. Interestingly, the neural network never incorrectly predicted class 3 and was nearly always correct in its predictions on data of class 3. In conclusion, the multi-input model performed best.

\begin{table}
\centering
\makebox[\textwidth]{
\begin{tabular}{c|c|c|c|c|c|c|c|c|c|}
\multicolumn{10}{c}{Predicted Class}\\
\cline{2-10}
\multirow{9}{*}{\rotatebox[origin=c]{90}{Real Class}} &  \cellcolor{lightgray} 0.9345 & 0.0036 & 0. & 0.0073 & 0.0036 & 0.0255 & 0. & 0.0254 & 0. \\
\cline{2-10}  & 0.0061 &  \cellcolor{lightgray} 0.9757 & 0. & 0. & 0. & 0.0061 & 0.0040 & 0.0061 & 0.0020 \\
\cline{2-10}  & 0. & 0. &  \cellcolor{lightgray} 0.9981 & 0. & 0. & 0.0019 & 0. & 0. & 0.         \\
\cline{2-10}  & 0. & 0.0118 & 0. &  \cellcolor{lightgray} 0.9294 & 0. & 0.0588 & 0. & 0. & 0.         \\
\cline{2-10}  & 0. & 0.1429 & 0. & 0.1429 &  \cellcolor{lightgray} 0.5714 & 0.1429 & 0. & 0. & 0.         \\
\cline{2-10}  & 0.0074 & 0.0222 & 0. & 0. & 0. &  \cellcolor{lightgray} 0.9630 & 0. & 0.00741 & 0.         \\
\cline{2-10}  & 0.0139 & 0.0694 & 0. & 0. & 0. & 0.0278 &  \cellcolor{lightgray} 0.8611 & 0.0278 & 0.            \\
\cline{2-10}  & 0.0390 & 0.0049 & 0. & 0.0098 & 0. & 0.0489 & 0. &  \cellcolor{lightgray} 0.8878 & 0.0098  \\
\cline{2-10}  & 0. & 0. & 0. & 0. & 0. & 0.0056 & 0.0056 & 0.0225 & \cellcolor{lightgray}  0.9663 \\
\cline{2-10}
\end{tabular}}
\caption{Multi-input algorithm: normalized confusion matrix for neural network with accuracy of 0.96. The balanced accuracy was 0.90 and the average table of confusion accuracy was 0.99 (not shown in the table).}
\label{table99}
\end{table}

\subsection{Ensemble}

The  ensemble algorithm performed second best of the four algorithms, slightly better than the text algorithm. The best neural network produced a best test accuracy of 0.95, a balanced accuracy of 0.82, and an average table of confusion  accuracy of 0.99. This used a text  neural network with a test accuracy of 0.94 and  a hex  neural network with a test accuracy of 0.91. This is shown in the second row of Table \ref{table78}.

\begin{table}
\centering
\begin{tabular}{|c|c||c|}
\hline
\rot{75}{text accuracy}  & \rot{75}{hex accuracy} & \rot{75}{max accuracy}  \\
\hline
\hline 0.87 &  0.91 & 0.93 \\
\hline 0.94 &  0.91 & \cellcolor{lightgray} 0.95 \\ 
\hline 0.90 &  0.91 & 0.93 \\
\hline 0.91 &  0.91 & 0.94 \\ 
\hline 0.90 &  0.92 & 0.94 \\
\hline 0.89 &  0.90 & 0.93 \\
\hline 0.89 &  0.88 & 0.92 \\
\hline 0.87 &  0.81 & 0.91 \\
\hline 0.89 &  0.82 & 0.91 \\
\hline
\end{tabular}
\caption{Ensemble algorithm: Each row shows the maximum test accuracy for ten runs. The best neural network from all the combinations of parameters has a test accuracy of 0.95. It used a text  neural network with a test accuracy of 0.94 and  a hex  neural network with a test accuracy of 0.91.}
\label{table78}
\end{table}

The normalized confusion matrix in Table \ref{table45} results from  the best neural network (test accuracy of 0.95).  The highlighted diagonal of the ensemble confusion matrix shows the proportions of predictions for each category that were correct. This ensemble neural network  predicted correctly above 83\% on all classes except class 5, which the neural network never predicted. This performance on class 5 is  similar to the best hex neural network mentioned above. In conclusion, the ensemble model performed second best, slightly better than the text model and worse than the multi-input model.

\begin{table}
\centering
\makebox[\textwidth]{%
\begin{tabular}{c|c|c|c|c|c|c|c|c|c|}
\multicolumn{10}{c}{Predicted Class}\\
\cline{2-10}
\multirow{9}{*}{\rotatebox[origin=c]{90}{Real Class}} &  \cellcolor{lightgray} 0.9273 & 0.0182 & 0.0073 & 0. & 0. & 0.0255 & 0. & 0.0145 & 0.0072 \\
\cline{2-10}  &  0.0040 & \cellcolor{lightgray} 0.9696 & 0. & 0. & 0. & 0.0061 & 0.0020 & 0.0142 & 0.0040 \\
\cline{2-10}  &  0.0019 & 0. & \cellcolor{lightgray} 0.9963 & 0. & 0. & 0.0019 & 0. & 0. & 0. \\
\cline{2-10}  &  0. & 0. & 0. & \cellcolor{lightgray} 0.9059 & 0. & 0.0706 & 0. & 0.0118 & 0.0118 \\
\cline{2-10}  &  0. & 0.2857 & 0. & 0.1429 & \cellcolor{lightgray} 0. & 0.4286 & 0. & 0. & 0.1429 \\
\cline{2-10}  &  0. & 0.0074 & 0.0074 & 0.0074 & 0. & \cellcolor{lightgray} 0.9704 & 0. & 0.0074 & 0. \\
\cline{2-10}  &  0.0694 & 0.0417 & 0. & 0. & 0. & 0.0278 & \cellcolor{lightgray} 0.8333 & 0.0278 & 0. \\
\cline{2-10}  &  0.0585 & 0.0049 & 0.0146 & 0.0146 & 0. & 0.0439 & 0.0049 & \cellcolor{lightgray} 0.8585 & 0. \\
\cline{2-10}  &  0.0056 & 0.0169 & 0. & 0.0112 & 0. & 0.0056 & 0. & 0.0169 & \cellcolor{lightgray} 0.9438 \\\cline{2-10}
\end{tabular}}
\caption{Ensemble algorithm: normalized confusion matrix for neural network with accuracy of 0.95. The balanced accuracy was 0.82 and the average table of confusion accuracy was 0.99 (not shown in the table).}
\label{table45}
\end{table}

\section{Conclusion}

\begin{figure}
\begin{center}
\includegraphics[width=1.0\textwidth]{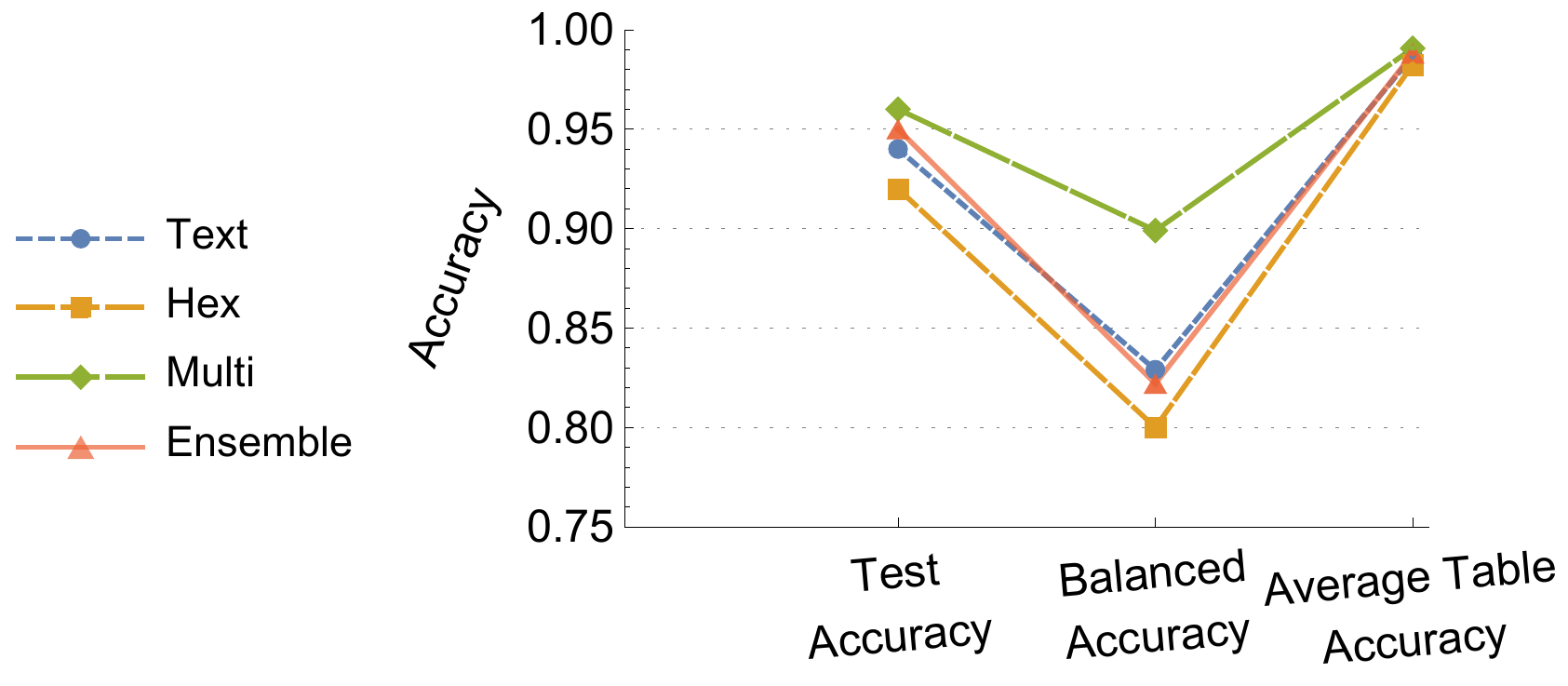}
\caption{For the test accuracy and the average table accuracy, we have \newline Multi-Input Accuracy $>$ Ensemble Accuracy $>$ Text Accuracy $>$ Hex Accuracy. For the balanced accuracy, we have \newline Multi-Input Accuracy $>$  Text Accuracy $>$ Ensemble Accuracy $>$ Hex Accuracy. The text and ensemble lines are quite close. \label{allAccuracies}}%
\end{center}
\end{figure}

For the test accuracy and the average table of confusion accuracy, our best multi-input model performed better than our best ensemble model, which performed slightly better than our best text model.  The hex model performed worst. For the balanced accuracy, the text and ensemble's performances are switched; the ensemble model's slightly worse performance may be due to its failure to correctly predict class 5. This is shown in Figure \ref{allAccuracies}. We thus think the multi-input algorithm may be most useful of the four algorithms for classifying this malware data.

The relatively poor performance of the hex data may be related to the fact there are around 256 possible hex commands because a hex command has two hexadecimal digits. This means more integer data (accomplishable using a larger max length) would be necessary to train a network to classify as well as the text model. This was not done because it would require more training time and we elected to keep the training under 24 hours.

The best ensemble and text models performed similarly  for all three types of accuracies. The similarity appears to be related to the pattern of lower accuracy from the hex data. Since the hex data is less useful, the network may learn to largely ignore the hex data and effectively perform a single-input investigation using the text data. That said, the ensemble model seems to have retained the ignorance of class 5 from the hex model.

\section{Future Work}

Using the same algorithms, to train a model that has better results we can optimize several parameters. If we increase the sequence length, which we recall is the quantity of integers retained from each file, the neural network could learn from more input and make more educated predictions. If we increase the vocabulary size, which we recall is the number of unique integers kept, the neural network may be able to pick up on more subtle details and predict better. We could also optimize various parameters of the layers in the training process, such as the batch size, which is the number of datapoints (files) processed simultaneously.

We could also create other algorithms  to train and compare their results to the text and multi-input algorithms. We could search for another way to parse the original asm files to get another format alongside the text and hex formats. We could also search for other ways to combine models, similar to the multi-input and ensemble methods.

Trying these four algorithms on different sets of malware would be informative. The dataset used here was specific to the Microsoft operating system. Would these algorithms rank the same in performance against other types of malware data sets?  And more generally, against other file-based data sets? Alternatively, perhaps static based analysis will become a thing of the past as real-time anomaly detection becomes the only way to effectively identify malware \emph{before} it does damage.

\section*{Acknowledgements}

This work was supported in part by the U.S. Department of Energy, Office of Science, Office of Workforce Development for Teachers and Scientists (WDTS) under the Science Undergraduate Laboratory Internship program.

This material is based upon work performed using computational resources supported by the University of Tennessee and Oak Ridge National Laboratory Joint Institute for Computational Sciences (http://www.jics.tennessee.edu).

\newpage

\bibliographystyle{plain}
\bibliography{../../../Bib/researchReferences1}

\begin{thebibliography}{1}

\bibitem{ChenLiMal}
Li~Chen.
\newblock Deep transfer learning for static malware classification.
\newblock {\em https://arxiv.org/abs/1812.07606}, 2018.

\bibitem{chollet2015keras}
Francois Chollet et~al.
\newblock Keras.
\newblock \url{https://keras.io}, 2015.

\bibitem{GoodallJohnR2019SIaE}
John~R Goodall, Eric~D Ragan, Chad~A Steed, Joel~W Reed, G.~David Richardson,
  Kelly~M.T Huffer, Robert~A Bridges, and Jason~A Laska.
\newblock Situ: Identifying and explaining suspicious behavior in networks.
\newblock {\em IEEE Transactions on Visualization and Computer Graphics},
  25(1):204--214, 2019.

\bibitem{lstm1997}
Sepp Hochreiter and J\"{u}rgen Schmidhuber.
\newblock Long short-term memory.
\newblock {\em Neural Comput.}, 9(8):1735--1780, November 1997.

\bibitem{McGuire:2007:GSP:1406599}
Paul McGuire.
\newblock {\em Getting Started with Pyparsing}.
\newblock O'Reilly, first edition, 2007.

\bibitem{RaffEtAl}
Edward Raff, Richard Zak, Russell Cox, Jared Sylvester, Paul Yacci, Rebecca
  Ward, Anna Tracy, Mark McLean, and Charles Nicholas.
\newblock An investigation of byte n-gram features for malware classification.
\newblock 2016.

\bibitem{microsoft}
Royl Ronen, Marian Radu, Corina Feuerstein, Elad Yom-Tov, and Mansour Ahmadi.
\newblock Microsoft malware classification challenge.
\newblock {\em http://arxiv.org/abs/1802.10135}, July 2018.

\bibitem{SewakMohit2018CoDL}
Mohit Sewak, Sanjay~K. Sahay, and Hemant Rathore.
\newblock Comparison of deep learning and the classical machine learning
  algorithm for the malware detection.
\newblock 2018.

\bibitem{ZhaoEtAl}
Hongwei Zhao, Mingzhao Li, Taiqi Wu, and Fei Yang.
\newblock Evaluation of supervised machine learning techniques for dynamic
  malware detection.
\newblock 2018.

\end{thebibliography}
\end{document}